# Acceleration of Solvation Free Energy Calculation via Thermodynamic Integration Coupled with Gaussian Process Regression and Improved Gelman-Rubin Convergence Diagnostics

Zhou Yu, Enrique R. Batista, Ping Yang, Danny Perez*

Theoretical Division, Los Alamos National Laboratory, Los Alamos, New Mexico 87545, United States

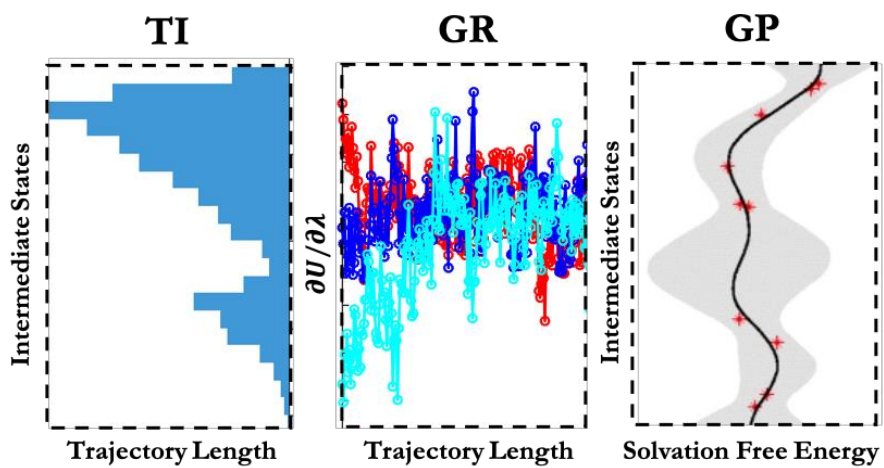

ORCID:
Zhou Yu:  0000-0003-3316-4979
Enrique R. Batista: 0000-0002-3074-4022
Ping Yang: 0000-0003-4726-2860
Danny Perez: 0000-0003-3028-5249

Corresponding Authors: Danny Perez (danny_perez@lanl.gov)

**Abstract:** The determination of the solvation free energy of ions and molecules holds profound importance across a spectrum of applications spanning chemistry, biology, energy storage, and the environment. Molecular dynamics simulations are a powerful tool for computing this critical parameter. Nevertheless, the accurate and efficient calculation of solvation free energy becomes a formidable endeavor when dealing with complex systems characterized by potent Coulombic interactions and sluggish ion dynamics and, consequently, slow transition across various metastable states. In the present study, we expose limitations stemming from the conventional calculation of the statistical inefficiency $g$ in the thermodynamic integration method, a factor that can hinder the determination of convergence of the solvation free energy and its associated uncertainty. Instead, we propose a robust scheme based on Gelman-Rubin convergence diagnostics. We leverage this improved estimation of uncertainties to introduce an innovative accelerated thermodynamic integration method based on Gaussian Process regression. This methodology is applied to the calculation of the solvation free energy of trivalent rare earth elements immersed in ionic liquids, a scenario where the aforementioned challenges render standard approaches ineffective. The proposed method proves effective in computing solvation free energy in situations where traditional thermodynamic integration methods fall short.

# 1. Introduction

The rare-earth elements (REEs) constitute a collection of seventeen chemically akin elements, encompassing fifteen lanthanides in addition to Scandium and Yttrium.[1] These elements are critical raw materials for a wide range of applications in modern technologies, especially in the areas of electronics, magnets, catalysis, and green energy technologies.[2] The pursuit of achieving a streamlined recycling process for REEs has garnered extensive attention, driven by the escalating demand and the challenges posed by supply chain dynamics.[3] Among the potential solution chemistry environments, ionic liquids (ILs) emerge as a promising contender because, composed of charged species, they can introduce strong perturbations in the electronic state of ions in solution, altering their chemical interaction properties. Characterized by their negligible vapor pressure, non-flammability, and exceptional chemical and electrochemical stability, ILs with solvent extractants offer a viable avenue to facilitate the efficient extraction of REEs from aqueous solutions.[4, 5] Nonetheless, the process of identifying effective species from a wide variety of ILs, or devising task-specific ILs, remains a time-intensive endeavor often plagued by laborious trial and error methodologies.[6]

Molecular dynamics (MD) simulation is, in principle, a powerful method for expediting the screening process to achieve specific relative coordination selectivity for particular elements in the mixture of dissolved ionic species through the calculation of solvation free energies.[7-9] These free energies delineate the thermodynamic cost and gain associated with the dissolution of a solute molecule within a solvent medium. Two prominent techniques, namely thermodynamic integration (TI) and free energy perturbation (FEP), find widespread applications in solvation free energy calculation.[10-12] In both methods, the $\lambda$ parameter smoothly transitions a system from an initial to a final state, such as from solute in vacuum ($\lambda=0$) to fully solvated ($\lambda=1$). This parameter facilitates the gradual modification of potential energy functions, enabling the estimation of free energy differences between intermediate states. TI and FEP differ in their approach toward calculating the free energy differences as a function of $\lambda$ parameters introduced in these methodologies. Specifically, the solvation free energy is calculated through the integration of the derivative of the Hamiltonian with respect to $\lambda$ between end states in the TI method, while solvation free energy is obtained through the summation of the free energy differences between neighboring states in the FEP method.

While the TI and FEP methods have broad applications in solvation free energy calculations across diverse systems, achieving robust convergence can be challenging, particularly in scenarios involving complex systems characterized by potent Coulombic interactions between solute and solvent and sluggish ion dynamics with slow transition across various metastable states. Ensuring that the simulations are adequately sampled and that the integration is performed with sufficient precision is crucial to obtaining accurate and reliable solvation free energy. In this work, we first take a system composed of $Eu^{3+}$ ion and ionic liquids 1-hexyl-3-methylimidazolium bis((perfluoroethyl)sulfonyl)imide (abbrev. [$C_6C_1Im$][BETI] see Figure 1a) as an example and discuss the limitations in the conventional methods used to estimate the statistical inefficiency *g*, which accounts for the correlation between samples and is crucial for determining the number of uncorrelated data points and solvation free energy uncertainties. Then, we propose an improved methodology based on the Gelman-Rubin (GR) convergence diagnostics[13] to determine the number of uncorrelated samples and more reliable solvation free energy uncertainties. Moreover, we propose an accelerated TI method by implementing Gaussian Process (GP) regression[14] and improved GR convergence diagnostics.[13] We underscore the hurdles associated with convergence during the computation of solvation free energy and highlight the potential of this novel algorithm in effectively capturing reliable solvation free energy and significantly enhancing the efficiency of solvation free energy computation. The subsequent sections are structured as follows: Section 2 delineates the simulation system, molecular model, and describes the methodologies, including TI methods, GR diagnostics, and GP regression; Section 3 discusses convergence challenges and the GP-augmented algorithm; conclusions are drawn in Section 4.

2. Methods

**Molecular systems:** For the MD trajectories exploring configuration space, we used a simulation cell consisting of 1 $Eu^{3+}$ ion, 600 [$C_6C_1Im$]$^+$ cations, and 603 [BETI]$^-$ anions, as shown in Figure 1, which ensures overall charge neutrality. Eu was chosen as a representative REE, and [$C_6C_1Im$][BETI] is a commonly used IL with bulky ionic structures. The initial configurations of the MD simulations were generated using the PACKMOL package.[15] The simulation box was periodic in all three dimensions. Relaxation under the NPT ensemble yielded a simulation box of approximately 7.3 nm on each side, which was deemed large enough to ensure that the perturbation on the distribution of ionic liquids induced by the $Eu^{3+}$ ion is negligible, as evidenced in the Eu-O radial distribution function shown in Figure S1 of the supporting information. The $Eu^{3+}$-ionic liquid and ionic liquid-ionic liquid interaction was modeled using classical force fields from recent publications,[16, 17] and the Lorentz-Berthelot mixing rule was used to obtain Lennard-Jones interaction parameters. All MD simulations were performed using the GROMACS 5.1.4 package.[18] Two distinct temperatures (e.g., 400 K and 600 K) underwent comprehensive investigation, both of which are below the decomposition temperature of the ionic liquid $[C_6C_1Im][BETI]$ (i.e., 638.8 K).[19]

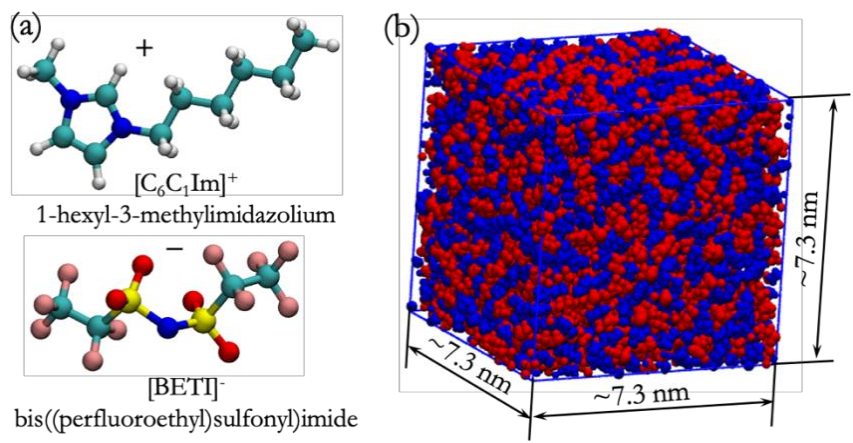

Figure 1. (a) Ionic liquid used in simulations. N, C, H, F, O, and S are denoted by cyan, blue, white, pink, red, and yellow spheres, respectively. (b) Snapshot of the MD simulation system, with anions and cations represented by red and blue spheres, respectively.

Because the initial configurations produced by PACKMOL tend to be high in energy, as PACKMOL only takes into account the geometry of the molecules but not the potential energy surface, a steepest-descent energy minimization was conducted, to quench the artificially introduced potential energy, until reaching a maximal force of 1000.0 kJ $mol^{-1}$ $nm^{-1}$. Subsequently, an equilibration run was performed under the NPT ensemble for 5 ns at the appropriate temperature, adopting a time step of 2 fs. The non-electrostatic interactions were computed via direct summation with a cut-off length of 1.2 nm. The electrostatic interactions were evaluated using the Particle Mesh Ewald (PME) method.[20] The real space cut-off and FFT spacing were 1.2 and 0.12 nm, respectively. All bonds with H-atoms were converted to rigid holonomic constraints using the LINCS algorithm to avoid the fast vibrational modes.[21] To maintain the system temperature at 400 K or 600 K, the Nose-Hoover thermostat[22, 23] with a time constant of 5 ps was employed. Simultaneously, the Parrinello-Rahman barostat[24, 25] maintained the system pressure at 1 atm with a time constant of 5 ps. The atomic coordinates and velocities were saved every 20 ps, and the system energies were saved every one ps.

**Thermodynamic Integration (TI) method:** The solvation free energy of the $Eu^{3+}$ ion within the ionic liquid was computed using the standard TI method. This involved modifying interactions between the $Eu^{3+}$ ion and ionic liquids, parameterized by the coupling parameter $\lambda_{LJ}$ and $\lambda_{Coulomb}$, spanning the range from 0 (representing a placeholder metal ion) to 1 (reflecting complete interaction). Subsequent MD simulations were conducted at multiple $\lambda$ values, connecting various intermediate states. As $[\lambda_{LJ}, \lambda_{Coulomb}]$ transitioned from [0, 0] to [1, 1], it delineated a path bridging the ionic liquids plus free $Eu^{3+}$ ion and the $Eu^{3+}$ ion in the ionic liquids system, enabling the calculation of free energy differences through the integration of the derivative of the free energy along this path. Specifically, to capture this transformation, a total of 26 intermediate states were chosen: $\lambda_{LJ}$ = 0, 0.2, 0.4, 0.6, 0.8, and 1.0 with $\lambda_{Coulomb}$= 0, signifying the gradual activation of LJ interactions. Subsequently, Coulombic interactions were incrementally introduced through $\lambda_{Coulomb}$ value of 0, 1/20, 2/20, ..., 19/20, and 1, while $\lambda_{LJ}$ remained fixed at 1. To ensure numerical stability during the LJ transformation, a soft-core potential was introduced to avoid singularities when $\lambda_{LJ}$ and $\lambda_{Coulomb}$ approached 0. The rationale of coupling parameter choices has been discussed in Figure S2 of the supporting information. For each intermediate state, a 5 ns NPT equilibration was performed, followed by a 30 ns production run at 600 K and a 50 ns production run at 400 K, respectively. The solvation free energy and its corresponding uncertainty were calculated through the established approach as follows:[10]

$$G_{i,i+1} = \frac{1}{2}w_i \left( \left\langle \left(\frac{\partial U}{\partial \lambda}\right)_i \right\rangle + \left\langle \left(\frac{\partial U}{\partial \lambda}\right)_{i+1} \right\rangle \right) \quad (1)$$

$$G = \sum_{i=1}^{K-1} G_{i,i+1} \quad (2)$$

$$\sigma_{i,i+1} = \frac{1}{2}\sqrt{w_i^2(\sigma_i^2 + \sigma_{i+1}^2)} \quad (3)$$

$$\sigma = \sqrt{\sum_{i=1}^{K-1} \sigma_{i,i+1}^2} \quad (4)$$

Where $G$, $\sigma$, $U$, and $K$ represent the solvation free energy, uncertainty, potential energy, and total number of intermediate states, respectively. $\sigma_i$ is the standard deviation of $\partial U/\partial \lambda$ at intermediate state $i$. Under the trapezoid rule, the weighting factors are $w_1 = w_K = 1/[2(K-1)]$ and $w_{k \neq 1,K} = 1/(K-1)$.

**Uncorrelated samples and statistical inefficiency:** The free energy expressions above should be applied to uncorrelated independent samples to estimate the statistical error or uncertainty in a stationary time series free of bias, which conventionally requires quantifying the effective number of uncorrelated samples and statistical inefficiency $g$ in the data set as follows:[10, 26]

$$\sigma^2_{i-equil} = \frac{\sigma_i^2}{N_{eff,i}} \quad (5)$$

$$N_{eff,i} = argmax\left(\frac{L_i - L_{i,0} + 1}{g_i}\right) \quad (6)$$

$$g_i = 1 + 2\tau_i \quad (7)$$

Argmax represents an argument that gives the maximum value from a target function. $L_i$ and $L_{i,0}$ denote the total number of samples and discarded samples maximizing number of uncorrelated samples (i.e., $N_{eff,i}$) in each intermediate state, respectively. The autocorrelation time $\tau_i$ in each intermediate state is defined as the integral of $\hat{\rho}_t(i)$, which is the autocorrelation function of observable $A$ (i.e., $\partial U/\partial \lambda$) and can be approximated in discretized form as follows:

$$\hat{\rho}_t(i) = \frac{\sum_{t=0}^{L} \delta A(t)\delta A(t+i)}{\sum_{t=0}^{L} \delta A^2(t)} \quad (8)$$

Here $\delta A(i)$ defines the deviation of the current observable from its empirical mean: $\delta A(i) = A(i) - \langle A(i) \rangle$.

**Improved Gelman-Rubin (GR) convergence diagnostics:** As will be demonstrated below, for the types of complex systems of interest here, the conventional error analysis described above can dramatically underestimate the true uncertainty of the free energy when the system can access multiple metastable states. In order to mitigate this issue, we have adopted an improved GR convergence diagnostics for convergence evaluation. In brief, the key idea behind the GR approach is that statistical inefficiency can be characterized by comparing the empirical means of a number of independent simulations (also called chains in this context) that are potentially initialized in different states. The comparison of the expectations measured by each chain then provides a convergence diagnostic. In practice, this method entails calculating the number of uncorrelated samples using the following expressions:[13]

$$\bar{A}_j = \frac{1}{L}\sum_{t=1}^{L} A_t^{(j)} \quad \text{Chain Mean} \quad (9)$$

$$\bar{A}_. = \frac{1}{J}\sum_{j=1}^{J} \bar{A}_j \quad \text{Grand Mean} \quad (10)$$

$$B = \frac{L}{J-1}\sum_{j=1}^{J}(\bar{A}_j - \bar{A}_.)^2 \qquad \text{Between Chain Variance} \qquad (11)$$

$$s_j^2 = \frac{1}{L-1}\sum_{t=1}^{L}(A_t^{(j)} - \bar{A}_j)^2 \qquad \text{Within Chain Variance} \qquad (12)$$

$$W = \frac{1}{J}\sum_{j=1}^{J} s_j^2 \qquad (13)$$

$$R = \sqrt{\frac{\frac{L-1}{L}W + \frac{1}{L}B}{W}} \qquad (14)$$

Here, $L$ represents the number of observables in each chain, and $J$ is the number of chains. One ps trajectory represents one observable and $J$ is 5 in this study. The number of uncorrelated samples for the combined chains is defined as:

$$S_{eff} = \frac{LJ}{\hat{\tau}} \qquad (15)$$

$$\hat{\tau} = 1 + 2\sum_{t=1}^{2k+1}\hat{\rho}_t = -1 + 2\sum_{t'=0}^{k}\hat{P}_{t'} \qquad (16)$$

$$\hat{P}_{t'} = \hat{\rho}_{2t'} + \hat{\rho}_{2t'+1} \qquad (17)$$

$$\hat{\rho}_t = 1 - \frac{W - \frac{1}{J}\sum_{j=1}^{J} s_j^2 \hat{\rho}_{t,j}}{\frac{L-1}{L}W + \frac{1}{L}B} \qquad (18)$$

Where $\hat{\rho}_{t,j}$ represents the autocorrelation of the observable for chain $j$. The initial positive sequence estimator is determined by selecting the largest $k$ for which $\hat{P}_{t'} > 0$ holds true for all $t' = 1, \ldots, k$. In this study, alongside simulations initialized and performed with a designated coupling parameter for each intermediate state, we performed four additional simulations at each value of λ, where each was initialized from the four closest neighboring intermediate states after a 5 ns NPT equilibration period. This strategy is designed to generate starting points with different structures of the solvation shell around the REE ion, limiting the probability that each one of them would be trapped in the same metastable state.

As will be shown below, the analysis of the errors contributed by each state (each value of λ) to the error on the total solvation free energy indicates that a large fraction is contributed by only a few states. In this case, carrying out the same amount of MD simulations in each state is highly inefficient. Instead, it would be preferable to invest the simulation time in the states with the highest statistical inefficiency so as to maximize the decrease of the overall uncertainty on the free energy with respect to the total simulation time aggregated over all states. To do so, we introduce two accelerated algorithms that allow for the identification of an optimal simulation schedule guided by the GR convergence diagnostics.

**Algorithm I** (Accelerated TI with improved GR convergence diagnostics): The schematic workflow of the proposed method is illustrated in Scheme A. First, we initiated the process by generating 5 short trajectories for each intermediate state (following the initialization scheme discussed above), from which we computed $\langle(\partial U/\partial\lambda)_i\rangle$ and $\sigma_i$ of the concatenated trajectories. Using the GR method, we then determined the number of uncorrelated samples $S_{eff,i}$ and the corresponding uncertainty $\sigma_i/\sqrt{S_{eff,i}}$, accounting for statistical insufficiency for each intermediate state. Then, the solvation free energy and its associated uncertainty were calculated following Equation 1-4 with the updated $\sigma_i/\sqrt{S_{eff,i}}$ for each intermediate state. The state for which an increase in trajectory length $\Delta L$ would reduce the overall uncertainty the most was then identified as follows. For each state, $S_{eff,i}$ is recalculated with an increased value of $L_i = L + \Delta L$ and $\hat{\tau}_i$ from last iterative loop, providing an estimation of the uncertainty on the total free energy if that specific trajectory was to be extended by $\Delta L$. The intermediate state leading to the largest reduction in $\sigma$ is then selected for additional simulation. This iterative loop repeats until either $\sigma$ or $L_i$ reaches a designated threshold.

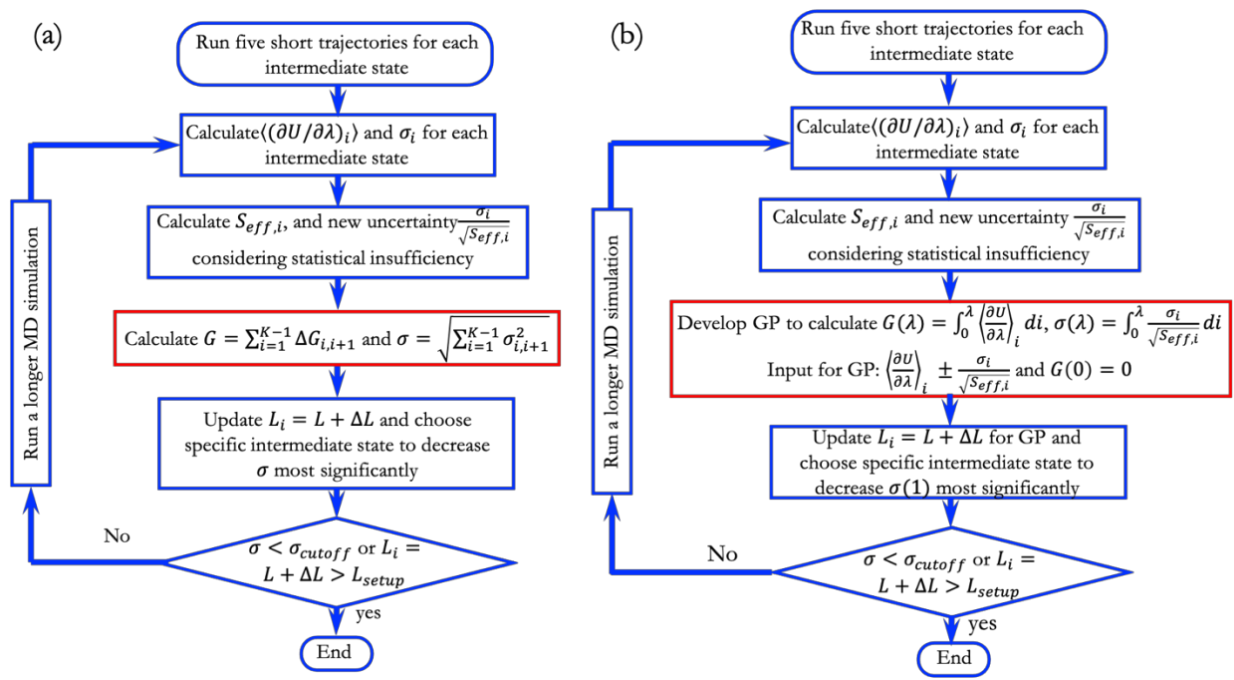

Scheme. Workflow of Algorithm (a) I and (b) II. The differences between these two algorithms are highlighted in the red boxes.

Algorithm I relies on the conventional TI approach, which is susceptible to discretization errors. Further, the method does not share information between the different states. For example, if the free-energy derivative is observed to vary smoothly with λ, a well-converged result at $\lambda_i$ should narrow the uncertainty at neighboring $\lambda_i$, hence potentially providing additional computational savings. These intuitive ideas can be formalized by representing the free energy profile as a Gaussian Process (GP), which essentially produces a posterior distribution of very general functions conditioned by discrete observations. GPs are one of the most popular methods in contemporary machine learning and so have been broadly reviewed elsewhere[27]. In contrast to conventional applications, the absolute value of the free energy is not accessible here, as only derivatives are measured at different values of λ. This information can, however, be naturally included in a GP to produce a posterior distribution of possible free energy profiles that are statistically consistent with the observed derivatives and their uncertainties[28].

**Algorithm II** (Accelerated TI with GP regression and improved GR convergence diagnostics)**:** The schematic workflow of the proposed algorithm is illustrated in Scheme B. Different from the previous method, the free energy and its associated uncertainty were determined utilizing the GP regression method, utilizing the point-wise free-energy derivatives $\langle(\partial U/\partial\lambda)_i\rangle$ and associated uncertainties $\sigma_i/\sqrt{S_{eff,i}}$ as input variables. Traditional regression models output a single deterministic prediction. In contrast, GP regression is a non-parametric Bayesian approach and treats the prediction as a probability distribution. In this study, the GP is characterized by a linear mean function representing the mean prediction at each point and a squared exponential kernel that defines the relationships between different points in the input space. The prefactor and the covariance length scale for the squared exponential kernel were constrained in the range [0, 10000] and [0.05 1.0], respectively. These hyper-parameters encode the smoothness properties of the function, i.e., the amplitude and characteristic length scales of the variations, respectively. The optimization of these hyperparameters is carried out through either Maximum a posteriori (MAP) or Markov chain Monte Carlo (MCMC) techniques with 300 chains and 2000 samples per chain. A burn-in phase discards the initial 100 samples from each chain. The implementation of GP with derivative constraints was achieved using the gptools package.[14] Furthermore, the GP regression was iteratively conducted, incorporating updated $\sigma_i/\sqrt{S_{eff,i}}$ as input for each intermediate state. As in Algorithm 1, $S_{eff,i}$ was recalculated for each state *i* with a candidate augmented time of $L_i =$ $L + \Delta L$ while utilizing $\hat{\tau}_i$ calculated from the last iterative loop. The specific intermediate state leading to the most substantial reduction in $\sigma$ was subjected to extended simulation duration. This iteration process stopped until either $\sigma$ or $L_i$ reached the predefined threshold.

## 3. Results and Discussion

*3.1 Challenges of solvation free energy calculations*

The assessment of uncertainty in solvation free energy calculations (Equations 3 to 4) applies only to a collection of uncorrelated energy derivative samples. This mandates a meticulous evaluation of the effective number of uncorrelated samples ($N_{eff}$) and the statistical inefficiency ($g$) in the datasets following Equations 6 to 7. A smaller $g$ corresponds to a greater abundance of uncorrelated samples, resulting in a proportional reduction in solvation free energy uncertainty and vice versa. Figure 2 shows the evolution of $g$ across each intermediate state for a free energy calculation at 600 K. Notably, in Figure 2a, states 1 to 6–representative of van der Waals contributions–exhibit low $g$ values that stabilize very rapidly. The maximal $g$ stands at 4.27 ps, as calculated using a 5 ns trajectory, which yields a large number of uncorrelated samples. Increasing the trajectory length then leads to a proportional and monotonic reduction in uncertainty, as shown in the error bars of Figure 3a.

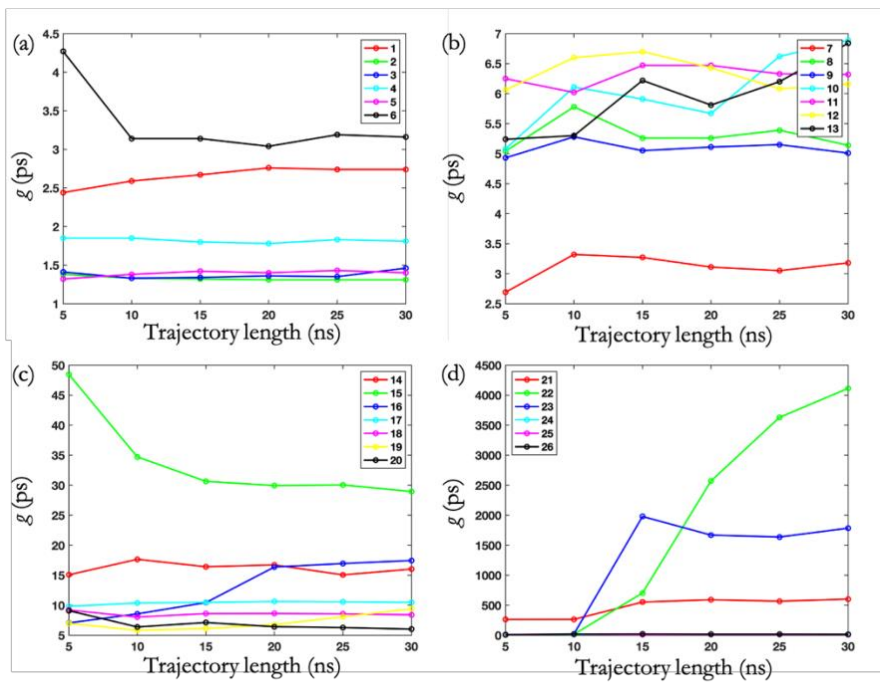

Figure 2. Evolution of statistical insufficiency $g$ maximizing $N_{eff}$ for each intermediate state at 600 K.

For the other intermediate states, within the initial 10 ns, the *g* values are also modest, as shown in Figures 2b-d, which leads to a seemingly converged solvation free energy associated with a very small uncertainty shown in Figure 3bc. However, a notable surge in *g* is observed beyond 10 ns for states 21, 22, and 23, reducing the estimated number of uncorrelated samples, which, consequently, leads to a dramatic increase in uncertainty, as shown in Figure 3bc.

These results underscore the danger of pseudo-convergence: in this case, relying on the effective number of uncorrelated samples ($N_{eff}$) and the statistical inefficiency (*g*) as the criteria for convergence could possibly lead to misjudgments on the calculation of solvation free energies and associated uncertainties. Our simulation exemplifies this pitfall; an ostensibly satisfactory solvation free energy and associated uncertainty was obtained from 10 ns trajectories, but further simulations indicate this assessment to be overly optimistic.

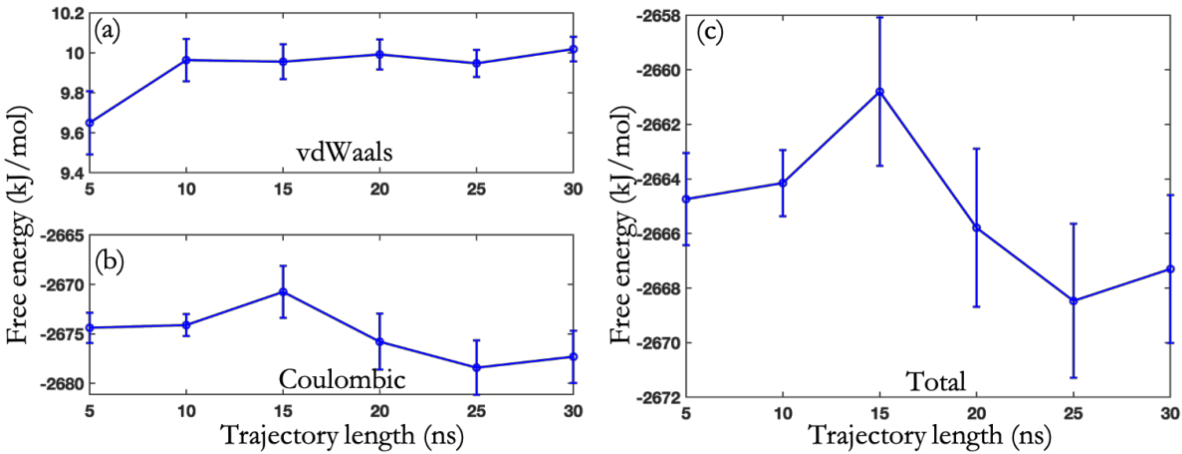

Figure 3. Evolution of (a) van der Waals contribution, (b) Coulombic contribution, and (c) overall solvation free energies, along with their associated uncertainties calculated via the thermodynamics integration (TI) method following Equations 1-8 at 600 K. $G_{i,i+1}$ and $\sigma_{i,i+1}$ used to calculate the solvation free energy and its uncertainty are summarized in Table S1 in the supporting information.

To rationalize the unexpected changes of *g* at intermediate states 21, 22, and 23, we report the corresponding evolution of $\partial U/\partial \lambda$ in Figure 4a. It can be seen that $\partial U/\partial \lambda$ exhibits a discontinuous behavior, with discrete jumps occurring on ~10 ns timescales. Within the initial 10 ns timeframe, the values of $\partial U/\partial \lambda$ for these three intermediate states are generally stable, which leads to a small estimated autocorrelation time and, subsequently, to a small *g*. However, as soon as a discrete transition in $\partial U/\partial \lambda$ occurs, the revised estimate of *g* increases significantly.

Further analysis shows that these fluctuations originate from the evolution of the solvation environment surrounding the Eu$^{3+}$ ion. The coordination number (CN) of oxygen atoms within [BETI]$^-$ ions, as well as the CN of [BETI]$^-$ ions in the first solvation shell surrounding the Eu$^{3+}$ ion in states 21, 22, and 23 are shown in Figure 4b-d, respectively. The number of atoms and ions surrounding the Eu$^{3+}$ ion were counted using a continuous CN function defined as $CN = \sum_{i=1}^{N} \frac{1-(r_i/r_0)^p}{1-(r_i/r_0)^q}$, with $p = 12$, $q = 24$, and the cutoff distance $r_0$ corresponds to the first valley in the radial distribution function from the Eu$^{3+}$ ion to oxygen atoms in [BETI]$^-$ ions (i.e., 0.34~0.36 nm) or center of mass of each [BETI]$^-$ ion (i.e., 0.64~0.66 nm), as explained in Figure S3. $r_i$ is the distance from the Eu$^{3+}$ ion to the $i^{\text{th}}$ O atom or [BETI]$^-$ ion.

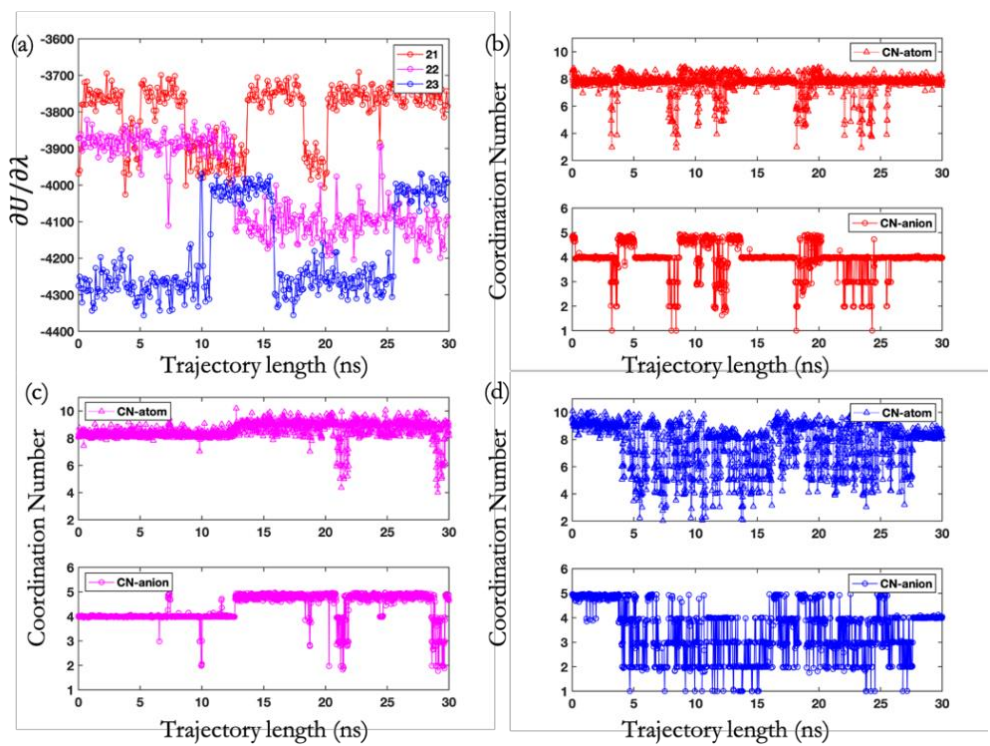

Figure 4. (a) Evolution of $\partial U/\partial \lambda$ for intermediate states 21, 22, and 23. (b-d) Variation in the coordination number of oxygen atoms and [BETI]$^-$ ions in the first solvation shell surrounding the Eu$^{3+}$ ion in intermediate states 21, 22, and 23, respectively, at 600 K.

Our analyses revealed diverse solvation environments featuring very sluggish solvation shell exchanges. For example, the Eu$^{3+}$ ion is coordinated by 4 [BETI]$^-$ ions for more than 12 ns and then coordinated by 5 [BETI]$^-$ ions in the intermediate state 22, as shown in Figure 4c. The changes observed in $\partial U/\partial \lambda$ exhibit a strong correlation with the variations in the CN of [BETI]$^-$ ions residing within the first solvation shell surrounding the Eu$^{3+}$ ion. Precisely, $\partial U/\partial \lambda$ follows an inverse relationship with the CN.

*3.2 Application of GR convergence diagnostics in TI methods*

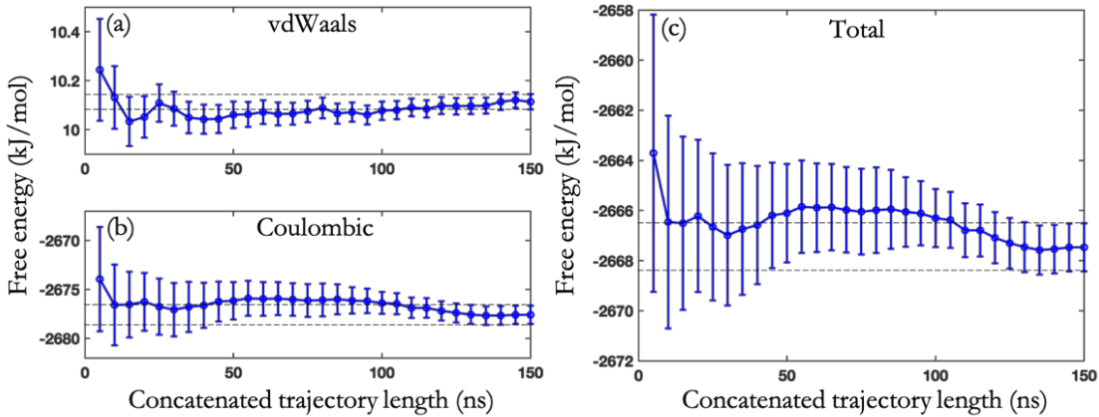

Figure 5. Evolution of (a) van der Waals contribution, (b) Coulombic contribution, and (c) overall solvation free energies, along with their associated uncertainties calculated via the thermodynamic integration (TI) method with the improved Gelman-Rubin (GR) convergence diagnostics at 600 K.

The above results indicate that the underestimation of the statistical inefficiency results from slow changes in the coordination environment of the ion. Unfortunately, the slower these changes occur, the more overly optimistic the assessment of the uncertainties, which is a very undesirable feature. However, GR statistics are, in principle, ideally suited to properly estimate the statistical inefficiency in such cases insofar as different trajectories sample different coordination environments. The observation that the preferred coordination environment changes with λ motivated our choice of the initialization procedure discussed above: by initializing an ensemble of simulations from initial states equilibrated at different values of λ, the probability of sampling from different metastable states is maximized, leading to a more accurate estimation of the uncertainties.

The corresponding solvation free energy is shown in Figure 5, where the time axis corresponds to the length of all concatenated trajectories–which corresponds to the true computational cost of the calculations. In this figure, the uncertainty computed from the concatenated trajectories is obtained using Equation 5, wherein $N_{eff}$ is replaced by $S_{eff}$ following the improved GR convergence diagnostics (see Equations 9 to 18).

Figure 5 shows a consistent and smooth reduction in solvation free energy uncertainty with increasing trajectory length at each intermediate state at 600 K. The improved GR convergence assessment avoids the emergence of the artificially underestimated uncertainty shown in Figures 3b and 3c, thereby providing a robustly converged free energy with its associated uncertainty. A concatenated trajectory length of 150 ns for each intermediate state yields converged free energy estimates: the van der Waals contribution, Coulombic contribution, and the overall solvation free energy stand at 10.1±0.0, -2677.6±0.9, and -2667.5±0.9 kJ/mol, respectively.

We also tracked the $S_{eff}$ changes across each intermediate state and observed a monotonic increase of $S_{eff}$ with the total trajectory length at 600 K. Particularly noteworthy is intermediate state 22, which again demonstrates the lowest $S_{eff}$; however, even in this instance, $S_{eff}$ remains above the recommended empirical convergence threshold of 50,[13] when the concatenated trajectory lengths exceed 115 ns, as shown in Figure S4. These findings suggest that solvation free energy calculations in systems with strong Coulombic interaction and sluggish ion dynamics greatly benefit from the improved GR convergence diagnostics strategy introduced above.

*3.3 Accelerated TI methods with GR convergence diagnostics and GP regression*

While the GR diagnostics improve the robustness of the results, the cost of these direct free energy calculations, where each state is assigned the same computational resources, proves extremely high. Using the accelerated TI workflows (i.e., Algorithm I & II), we undertook a reevaluation of the Coulombic contribution to solvation free energy and its associated uncertainty at 600 K. Specifically, initially, 5 replicas of 2 ns trajectory were generated for each intermediate state. Four out of these 5 trajectories were generated starting from the 4 nearest neighboring intermediate states following a 5 ns NPT equilibration period. Then, $S_{eff,i}$ and $\sigma_i/\sqrt{S_{eff,i}}$ were calculated according to the improved GR convergence diagnostics using these 5 replicas for each intermediate state. Subsequently, leveraging these values, we either utilized the TI method (i.e., Algorithm I see the red box in Scheme A) or employed a GP model (i.e., Algorithm II see the red box in Scheme B) to determine the optimal intermediate state requiring extended simulation trajectories to produce the largest expected reduction in the total free energy uncertainty. This chosen intermediate state, including all five replicas, was then subjected to an additional 2 ns extended simulation. This iterative process was repeated until the trajectory length of an intermediate state surpassed 30 ns, yielding a concatenated trajectory exceeding 150 ns. Throughout the iteration procedure involving GP regression in Algorithm II, we did not observe overfitting, and hyperparameters for the squared exponential kernel remained well within the designated range, far from the boundary, as shown in Figure S5.

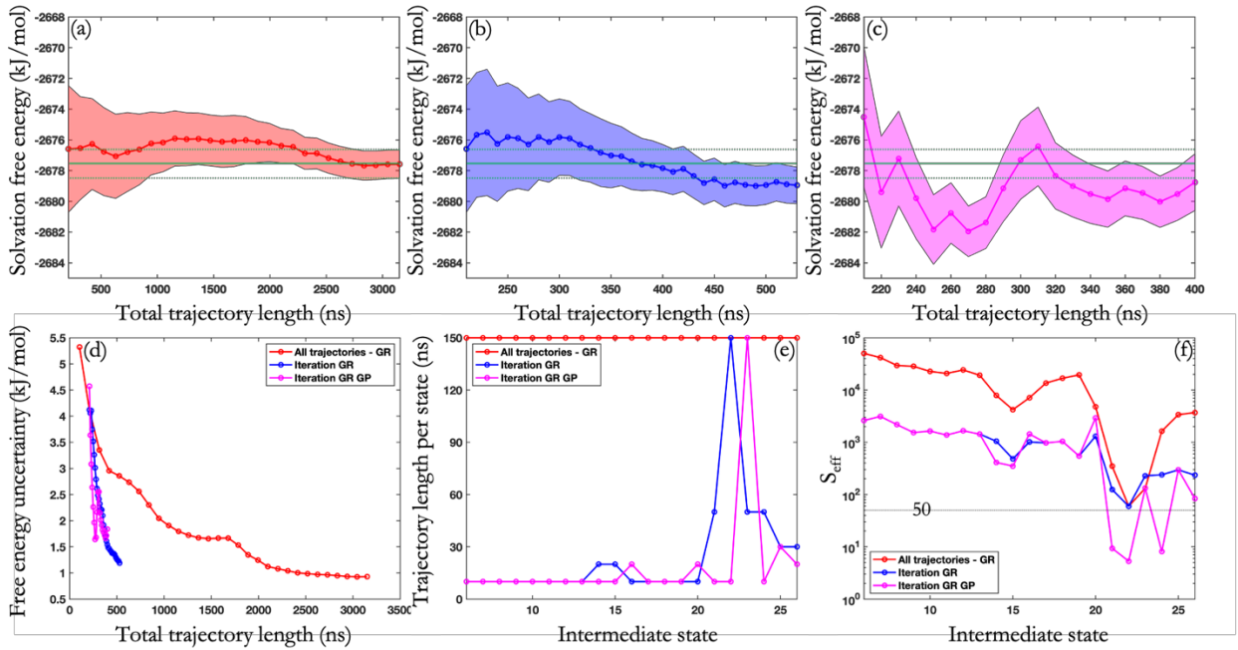

Figure 6. Evolution of Coulombic contribution to solvation free energy at 600 K using (a) TI method with GR diagnostics by uniformly expanding trajectories at each intermediate state, (b) Algorithm I (accelerated TI with GR diagnostics), and (c) Algorithm II (accelerated TI with GP regression-MCMC sampler and GR diagnostics). (d) Evolution of solvation free energy uncertainties corresponding to (a-c). (e) Trajectory length and (f) $S_{eff}$ of each intermediate state at the end of the calculation corresponding to (a-c). The shading areas in (a-c) represent the uncertainties. The solid green line accompanied by two dotted green lines in (a-c) illustrates the Coulombic part of solvation free energy and its uncertainty calculated using the TI method and GR diagnostics with a concatenated trajectory length of 150 ns for each intermediate state (as shown in the last point in (a)). The black dotted line in (f) signifies the empirical threshold of $S_{eff}$ for convergence.

From Figure 6, we can see that the newly developed algorithms have excellent performances in terms of both accuracy and efficiency. Specifically, we can see the eventually stabilized energy and its associated uncertainty calculated according to Algorithm I & II is consistent with the converged data derived from the TI method with GR diagnostics using a concatenated 150 ns trajectory for each intermediate state (c.f., solid and dotted green lines in Figure 6bc). From Figure 6d, we can see a rapid decrease in uncertainty over successive iterations following Algorithm I &

II. Eventually, the uncertainties from these two newly developed algorithms are also consistent with the conventional TI method, with GR diagnostics relying on considerably lengthier trajectories. The computational cost ($\eta$) of the accelerated TI method (Algorithm I & II), relative to the conventional TI methods with GR diagnostic, can be quantified as:

$$\eta = \frac{L_0 JK + \Delta L JI}{LK} \quad (19)$$

Here, $L_0$ and $\Delta L$ represent the initial trajectory length per each replica and intermediate state, and the timestep for the iteration process, respectively. They are both 2 ns in this context. $J$ presents the number of replicas per intermediate state (5 in this context) and $K$ denotes the number of intermediate states (21 for Coulombic contribution). $I$ represent the iteration counts and $L$ denotes the concatenated trajectory length per intermediate state in the conventional TI with GR diagnostics. By substituting $I = 33$ or 20 iterations and $L = 150$ ns, the above equation indicates that the computational cost diminishes to 17% or 13% following Algorithm I or II, respectively, compared to the TI method with GR diagnostics relying on 150 ns trajectory for each intermediate state.

The trajectory length and $S_{eff}$ for each intermediate state throughout the iterative process were tracked in the newly developed algorithms. For Algorithm I, it was found that the intermediate state with lower $S_{eff}$ consistently had more trajectories added to reduce the uncertainty in the solvation free energy. Eventually, $S_{eff}$ for all intermediate states exceeded the recommended threshold of 50, as shown in Figure 6f. For Algorithm II, three intermediate states (e.g., 21, 22, and 24) retain $S_{eff}$ below the recommended value 50 at the end of the iterations, as illustrated in Figure 6f. This can be explained by the fact that the GP model exploits the smoothness of the free energy with respect to λ in order to limit the amount of direct simulation required in states that are most difficult to converge. Indeed, the covariance length scale for the squared exponential kernel in the GP model shown in Figure S5b (which is automatically re-adjusted during the calculation) spans the range of 0.29~0.36, which is significantly larger than the coupling factor differences between two adjacent neighboring states. In other words, the GP leveraged the fact that the variations in free energy are smooth on the scale of the separation between adjacent states (which is unsurprising, given that this is an implicit requirement for TI to be accurate). Therefore, the GP model manifests a tendency to selectively disregard the substantial uncertainty from these specific points, ultimately yielding reliable energy estimates accompanied by reduced uncertainty. Note that we have found that a conventional Maximum a Posteriori (MAP) estimation of the hyperparameters tended to overestimate the smoothness of the function and, hence, underestimate uncertainties. We therefore recommend the use of Markov chain Monte Carlo (MCMC) to determine hyperparameter values. Nevertheless, MAP ultimately converged to a reliable estimate of solvation free energy, as illustrated in Figure S6. Additionally, the corresponding analysis of van der Waals contribution to solvation free energy at 600 K is presented in Figure S7 in the supporting information.

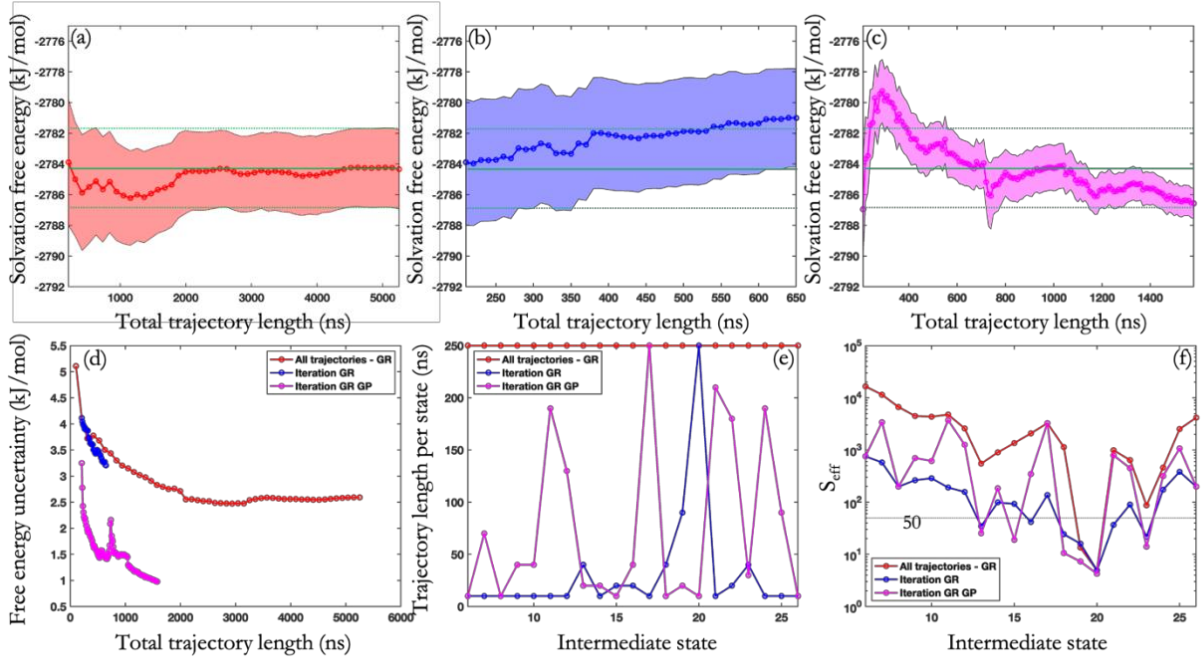

Figure 7. Evolution of Coulombic contribution to solvation free energy at 400 K using (a) TI method with GR diagnostics by uniformly expanding trajectories at each intermediate state, (b) Algorithm I (accelerated TI with GR diagnostics), and (c) Algorithm II (accelerated TI with GP regression-MCMC samplers and GR diagnostics). (d) Evolution of solvation free energy uncertainties corresponding to (a-c). (e) Trajectory length and (f) $S_{eff}$ for each intermediate state at the end of the calculation corresponding to (a-c).

In order to assess the performance of our newly developed algorithms under more challenging conditions, we conducted a comparative study at a lower temperature of 400 K. The lowered temperature slows down the transition among metastable states and further impedes the convergence of MD simulations. Firstly, we calculated the solvation free energy using the TI method with GR diagnostics, employing a 250 ns concatenated trajectory for each intermediate state, as depicted in Figure 7a. The uncertainty of solvation free energy gradually decreased with the increase in trajectory length. However, it remained significantly larger compared to the simulations at 600 K, as indicated by the red curves in Figures 6d and 7d. This discrepancy is primarily attributed to the smaller values of $S_{eff}$ observed at 400 K when compared to those at 600 K. Notably, $S_{eff,20}$ was found to be ~5, which is much lower than the recommended threshold of 50, as illustrated in Figure 7f. Furthermore, we examined the evolution of $\partial U/\partial \lambda$ among the 5 replicas within this specific intermediate state, shown in Figure 8a. One of the replicas exhibited distinctive behavior, characterized by a diminished presence of [BETI]$^-$ ions in the first solvation shell surrounding the Eu$^{3+}$ ion shown in Figure 8b, but transitions between different solvation states were observed to be very slow. This (correctly) leads to a very slow decrease in the uncertainties with additional simulation time. This would be a worst-case scenario of the conventional approach to estimate the conventional inefficiency, as pseudo-convergence would be rapidly established, severely underestimating the actual uncertainties.

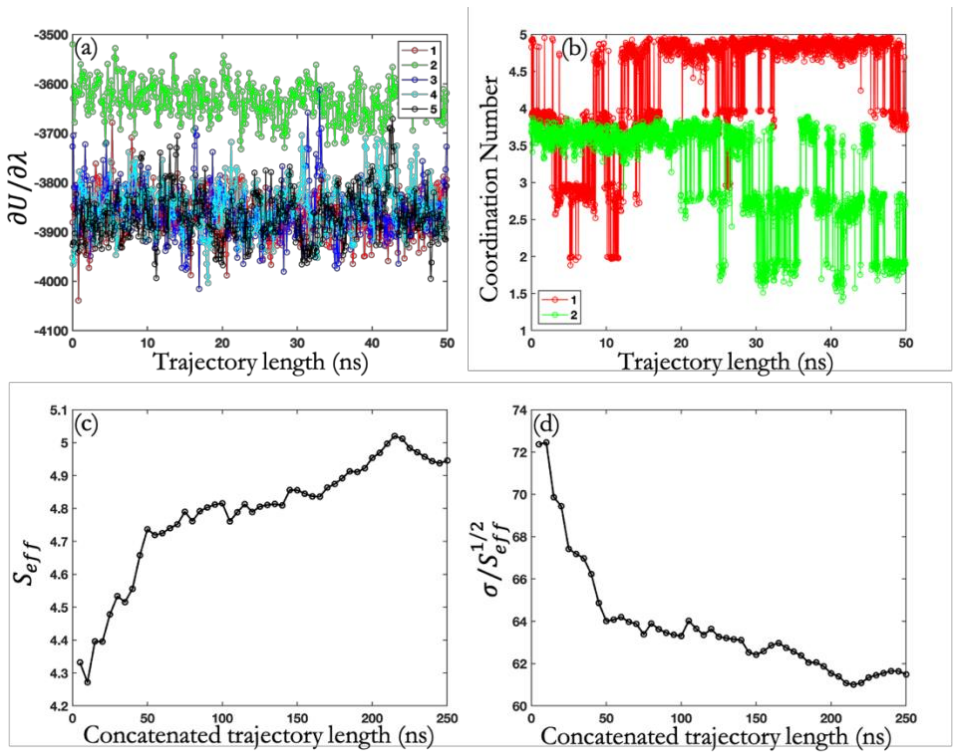

Figure 8. (a) Evolution of $\partial U/\partial \lambda$ among 5 replicas in intermediate state 20 at 400 K. (b) Variation in the coordination number of [BETI]$^-$ ions in the first solvation shell surrounding the Eu$^{3+}$ ion in replica 1 and 2 in intermediate state 20 at 400 K. Evolution of (c) $S_{eff}$ and (d) $\sigma/S_{eff}^{1/2}$ in intermediate state 20.

Subsequently, we reevaluated the solvation free energy and its uncertainty at 400 K using the newly developed Algorithm I & II. Similarly, the procedure started by generating 5 replicas of 2 ns trajectories for each intermediate state. The selected intermediate state, incorporating all five replicas, was then subjected to an additional 2 ns extended simulation. This iterative procedure persisted until the trajectory length of an intermediate state exceeded 50 ns, resulting in a concatenated trajectory exceeding 250 ns. Algorithm I, during the iterative process, demonstrated a propensity for consistently assigning more trajectories to the intermediate state 20 with the smallest $S_{eff}$ and largest $\sigma/S_{eff}^{1/2}$. However, $S_{eff,20}$ and $\sigma_{20}/S_{eff,20}^{1/2}$ exhibited relative stability due to the very low transition rates between metastable states, prohibiting the decrease of the uncertainty with additional simulation time, as shown in Figure 8cd. For Algorithm II, the simulation attempted to circumvent some difficult intermediate states, as explained above. Therefore, the uncertainty is much smaller, as shown in Figure 7d. For example, the iterative process in Algorithm II was not burdened with the need to extend the trajectory length of the intermediate state 20. When the overall trajectory length is shorter than 370 ns, we observed 9 intermediate states with $S_{eff}$ below 50. As the trajectory length increased, $S_{eff}$ demonstrated a tendency to rise. Ultimately, the stable solvation free energy with uncertainty resided within the range of solvation free energy and its associated uncertainty computed using all available trajectories, although 6 intermediate states are with $S_{eff}$ lower than 50. This indicates that these results should likely be viewed as optimistic since the estimates of the uncertainties in different states are likely not fully robust, leading to the slight (1.5-2x) underestimation of the actual uncertainties that can be inferred from Figure 7c.

These results show that the two proposed accelerated TI algorithms can be powerful tools to improve computational efficiency and facilitate convergence, offering a robust solution for intricate systems needing precise solvation free energy calculations. Algorithm I is straightforward and has a good performance in reducing computational cost. However, this algorithm may encounter challenges with intermediate states characterized by extremely slow transitions between metastable states. The incorporation of GP regression within Algorithm II can help mitigate challenging intermediate states with substantial uncertainty, ultimately facilitating convergence when the covariance length scale in the GP model exceeds the coupling factor differences between these intermediate states. Note that these improved algorithms singularly rely on correct estimates of the uncertainties at each state, as spurious underestimates of the type observed with the conventional method would be self-reinforcing, i.e., would discourage further evaluations in this state, perpetuating the uncertainty underestimation.

## 4. Conclusions

In this study, we have tackled the challenge of accurately and efficiently computing solvation free energies for systems characterized by strong Coulombic interactions and sluggish ion dynamics, such as trivalent rare earth elements immersed within ionic liquids. Our investigation highlighted the limitations associated with the statistical inefficiency $g$ and the number of uncorrelated samples (i.e., $N_{eff,i}$) in the conventional TI method, an impediment to robustly determining convergence and associated uncertainties in solvation free energy. Our findings demonstrate that the convergence of solvation free energy calculations is influenced by the coordination dynamics of solute-solvent interactions, which can lead to deceptive convergence indications. To address this, we have demonstrated the usefulness of an improved GR convergence diagnostics approach. Furthermore, we have introduced two innovative accelerated TI algorithms that incorporate GP regression and improved GR convergence diagnostics, which ensure robust convergence by adapting trajectory lengths and refining uncertainties through iterative optimization. These methodologies show promise in accurately capturing solvation free energy while substantially reducing computational costs compared to traditional methods, potentially opening the door to reliable high-throughput simulations of solvation free energy calculations.

**Supporting Information**

The Supporting Information is available free of charge at https://pubs.acs.org/doi/***

Radial distribution function from $Eu^{3+}$ ion to oxygen atoms in $[BFTI]^-$ ions; Intermediate states choices; Radial distribution function from $Eu^{3+}$ ion to oxygen atoms in $[BFTI]^-$ ions in intermediate states 21, 22, and 23; Evolution of $S_{eff}$ for the intermediate 22; Evolution of prefactor and covariance length scale for the squared exponential kernel in the GP model; Evolution of van der Waals and Coulombic contribution to the solvation free energy and corresponding uncertainty at 400 K and 600 K.

**Author information**

**Corresponding author:**

Danny Perez - Theoretical Division, Los Alamos National Laboratory, Los Alamos, New Mexico 87545, United States; orcid.org/0000-0003-3028-5249

**Authors:**

Zhou Yu - Theoretical Division, Los Alamos National Laboratory, Los Alamos, New Mexico 87545, United States; orcid.org/0000-0003-3316-4979

Ping Yang - Theoretical Division, Los Alamos National Laboratory, Los Alamos, New Mexico 87545, United States; orcid.org/0000-0003-4726-2860

Enrique R. Batista - Theoretical Division, Los Alamos National Laboratory, Los Alamos, New Mexico 87545, United States; orcid.org/0000-0002-3074-4022

**Notes:** The authors declare no competing financial interest.

**Acknowledgement:** This material is based upon work supported by the U.S. Department of Energy, Office of Science, Office of Basic Energy Science, under Award Number KC0302020. LANL is operated by Triad National Security, LLC, for the National Nuclear Security Administration of U.S. Department of Energy (contract no. 89233218CNA000001).

# Supporting Information

Acceleration of Solvation Free Energy Calculation via Thermodynamic Integration Coupled with Gaussian Process Regression and Improved Gelman-Rubin Convergence Diagnostics

Zhou Yu, Enrique R. Batista, Ping Yang, Danny Perez*

Theoretical Division, Los Alamos National Laboratory, Los Alamos, New Mexico 87545, United States

ORCID:
Zhou Yu:  0000-0003-3316-4979
Enrique R. Batista: 0000-0002-3074-4022
Ping Yang: 0000-0003-4726-2860
Danny Perez: 0000-0003-3028-5249

Corresponding Authors: Danny Perez (danny_perez@lanl.gov)

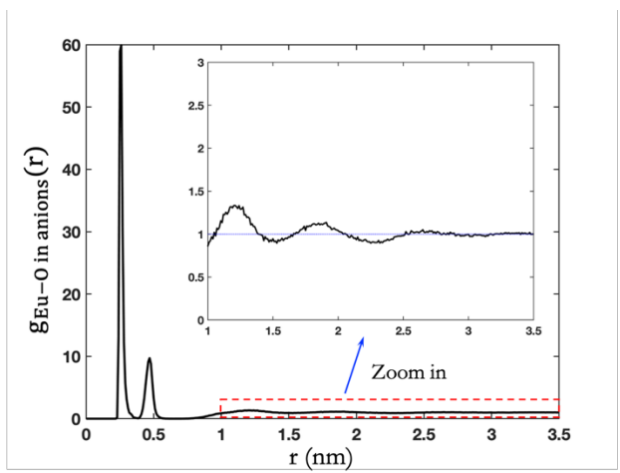

Figure S1. Radial distribution function from the $Eu^{3+}$ ion to the oxygen atoms in [BFTI]$^-$ ions.

Figure S1 shows the radial distribution function from the $Eu^{3+}$ ion to the oxygen atom in ionic liquid anions stabilizes at 1 when the distance is larger than ~3 nm. This means the box size (~7.3 nm) after volume relaxation under the NPT ensemble in this work is reasonable to neglect the interference on the distribution of ionic liquids induced by the mirrored $Eu^{3+}$ ion due to periodic boundary conditions.

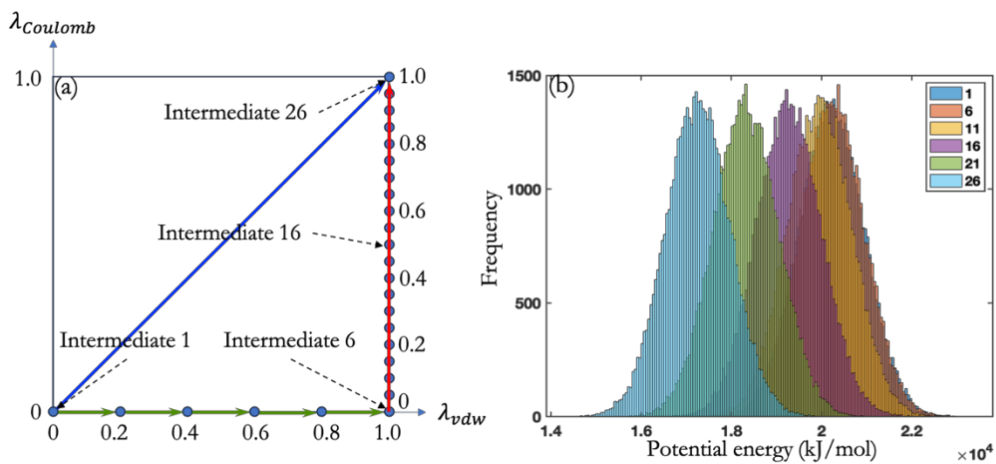

Figure S2. (a) The intermediate states pathway is shown as a Cartesian plane formed by the axes $\lambda_{LJ}$ and $\lambda_{Coulomb}$, which control van der Waals and Coulombic interactions between $Eu^{3+}$ ion and ionic liquids. Specifically, $\lambda = [0, 0]$ (intermediate state 1) means the $Eu^{3+}$ ion is no longer coupled to the system; $\lambda = [1.0, 0]$ (intermediate state 6) corresponds to the electrostatically non-interacting molecule; and $\lambda = [1.0, 1.0]$ (intermediate 26) means the interaction between $Eu^{3+}$ ion and ionic liquids is fully considered. The van der Waals (x-axis) and Coulombic (y-axis) interactions are adjusted by 6 and 21 evenly distributed lambda points, respectively. (b) The histograms of potential energies of intermediate state 1, 6, 11, 16, 21, and 26.

Figure S2b shows significant overlaps on potential energy histograms in various intermediate states, which implies that the choice of intermediate states/coupling factors is reasonable.

Table S1 $G_{i,i+1}$ and $\sigma_{i,i+1}$ calculated with different trajectory lengths

| States | 5 ns (kJ/mol) | 10 ns (kJ/mol) | 15 ns (kJ/mol) | 20 ns (kJ/mol) | 25 ns (kJ/mol) | 30 ns (kJ/mol) |
|---|---|---|---|---|---|---|
| 0 - 1 | -1.147+-0.005 | -1.150+-0.003 | -1.145+-0.003 | -1.146+-0.002 | -1.145+-0.002 | -1.146+-0.002 |
| 1 - 2 | 0.028+-0.014 | 0.025+-0.010 | 0.034+-0.008 | 0.041+-0.007 | 0.036+-0.006 | 0.036+-0.006 |
| 2 - 3 | 2.843+-0.054 | 2.850+-0.038 | 2.841+-0.031 | 2.858+-0.027 | 2.835+-0.024 | 2.857+-0.022 |
| 3 - 4 | 4.584+-0.088 | 4.711+-0.063 | 4.704+-0.052 | 4.722+-0.045 | 4.707+-0.041 | 4.733+-0.037 |
| 4 - 5 | 3.342+-0.118 | 3.526+-0.075 | 3.521+-0.062 | 3.517+-0.054 | 3.515+-0.048 | 3.537+-0.044 |
| 5 - 6 | -18.716+-0.139 | -18.789+-0.095 | -18.779+-0.077 | -18.631+-0.065 | -18.661+-0.059 | -18.697+-0.054 |
| 6 - 7 | -30.237+-0.154 | -30.179+-0.118 | -30.294+-0.093 | -30.172+-0.080 | -30.187+-0.071 | -30.170+-0.065 |
| 7 - 8 | -43.023+-0.180 | -42.980+-0.134 | -43.046+-0.106 | -43.023+-0.091 | -43.023+-0.082 | -43.009+-0.074 |
| 8 - 9 | -56.130+-0.181 | -56.252+-0.138 | -56.216+-0.111 | -56.280+-0.095 | -56.331+-0.089 | -56.163+-0.081 |
| 9 - 10 | -69.810+-0.195 | -69.807+-0.144 | -69.688+-0.119 | -69.643+-0.102 | -69.722+-0.094 | -69.567+-0.086 |
| 10 - 11 | -83.394+-0.205 | -83.452+-0.146 | -83.243+-0.121 | -83.259+-0.104 | -83.210+-0.091 | -83.218+-0.083 |
| 11 - 12 | -96.333+-0.190 | -96.278+-0.137 | -96.182+-0.117 | -96.301+-0.098 | -96.273+-0.088 | -96.374+-0.083 |
| 12 - 13 | -109.407+-0.253 | -108.857+-0.189 | -108.734+-0.155 | -108.768+-0.134 | -109.031+-0.117 | -109.266+-0.111 |
| 13 - 14 | -122.378+-0.461 | -122.295+-0.298 | -122.259+-0.232 | -122.543+-0.200 | -122.812+-0.176 | -122.779+-0.161 |
| 14 - 15 | -135.911+-0.429 | -136.307+-0.270 | -136.385+-0.215 | -136.840+-0.197 | -136.614+-0.177 | -136.481+-0.161 |
| 15 - 16 | -148.949+-0.218 | -148.919+-0.161 | -148.844+-0.140 | -149.011+-0.139 | -148.822+-0.127 | -148.888+-0.116 |
| 16 - 17 | -159.645+-0.222 | -159.371+-0.153 | -159.490+-0.128 | -159.426+-0.111 | -159.621+-0.098 | -159.549+-0.089 |
| 17 - 18 | -168.670+-0.189 | -168.454+-0.124 | -168.764+-0.104 | -168.931+-0.092 | -169.054+-0.085 | -168.922+-0.080 |
| 18 - 19 | -177.068+-0.174 | -176.835+-0.107 | -176.905+-0.092 | -177.131+-0.079 | -177.110+-0.074 | -177.042+-0.070 |
| 19 - 20 | -186.365+-0.824 | -186.031+-0.588 | -186.045+-0.713 | -185.823+-0.651 | -185.609+-0.555 | -185.132+-0.515 |
| 20 - 21 | -192.696+-0.818 | -192.879+-0.595 | -194.129+-1.039 | -195.043+-1.552 | -195.997+-1.640 | -195.357+-1.539 |
| 21 - 22 | -203.723+-0.165 | -204.096+-0.187 | -202.255+-1.677 | -204.287+-1.854 | -205.652+-1.841 | -205.652+-1.777 |
| 22 - 23 | -216.506+-0.214 | -216.977+-0.185 | -213.941+-1.498 | -214.961+-1.207 | -215.071+-1.007 | -215.333+-1.029 |
| 23 - 24 | -223.901+-0.217 | -224.130+-0.173 | -224.114+-0.124 | -224.301+-0.116 | -224.147+-0.104 | -224.224+-0.090 |
| 24 - 25 | -231.530+-0.180 | -231.227+-0.177 | -231.441+-0.132 | -231.407+-0.115 | -231.469+-0.104 | -231.494+-0.086 |

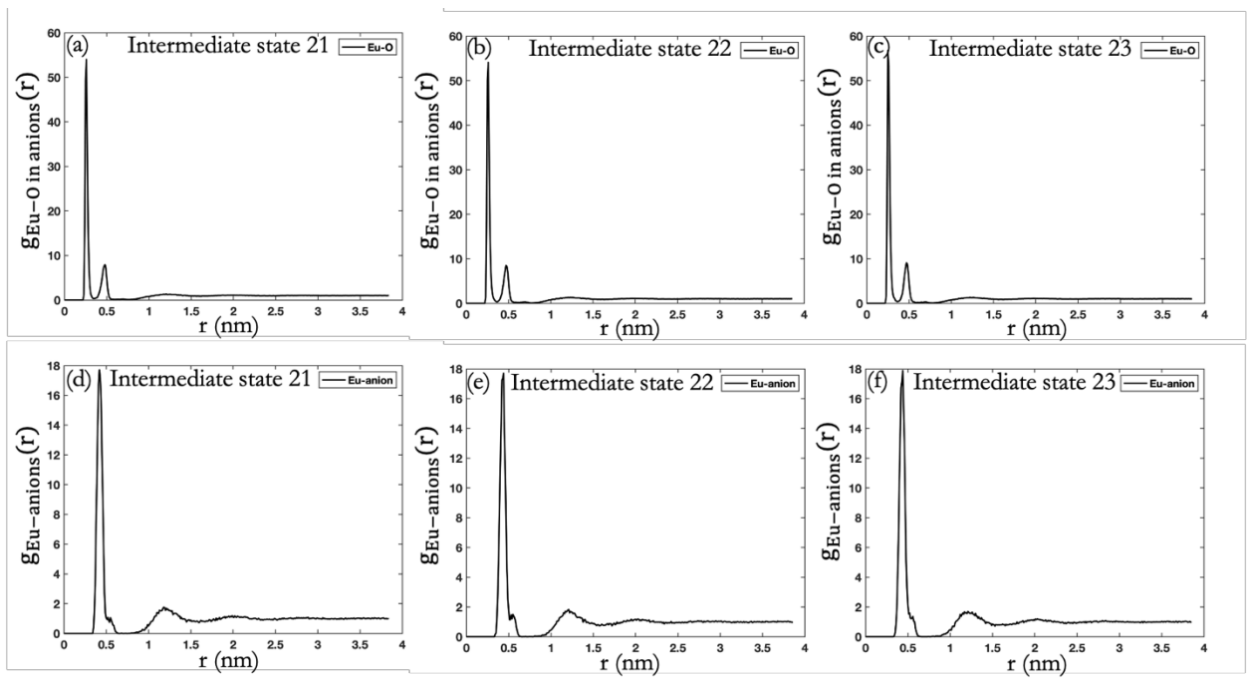

Figure S3. Radial distribution function from the $Eu^{3+}$ ion to the oxygen atoms in $[BFTI]^-$ ions or center of mass of each $[BFTI]^-$ ion in intermediate states (a, d) 21, (b, e) 22, and (c, f) 23, respectively. The first valley locations of RDFs (i.e., cutoff distance $r_0$) in (a-f) are 0.34, 0.36, 0.36, 0.64, 0.65, and 0.66 nm, respectively.

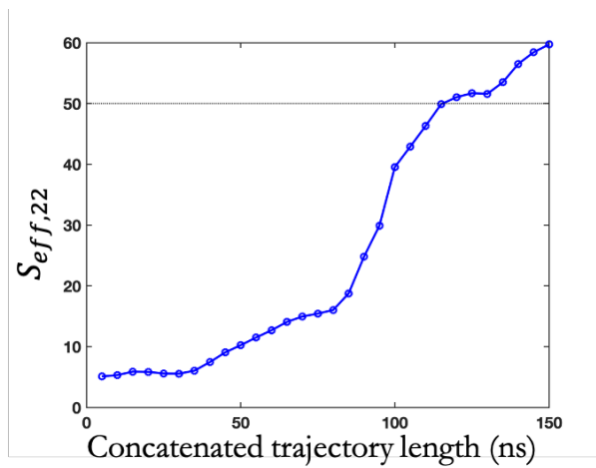

Figure S4. Evolution of $S_{eff}$ for the intermediate state 22 at 600 K.

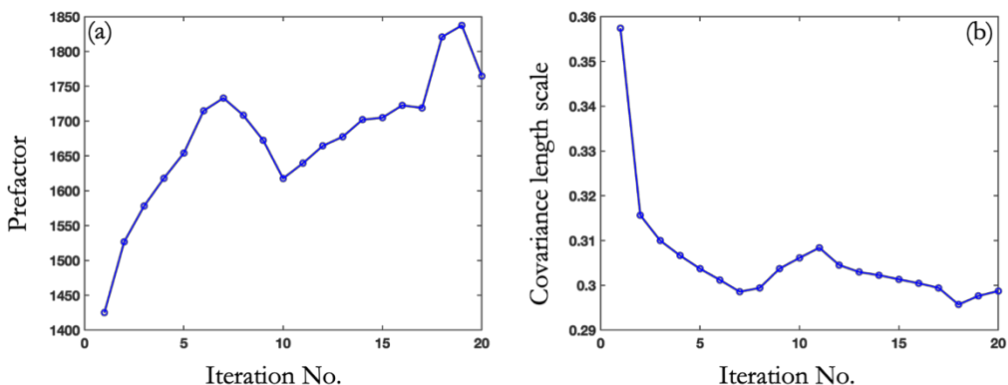

Figure S5. Evolution of prefactor and covariance length scale for the squared exponential kernel in the GP model with MCMC. The setup boundary for the prefactor and covariance length scale is [0, 10000] and [0.05 1.0], respectively.

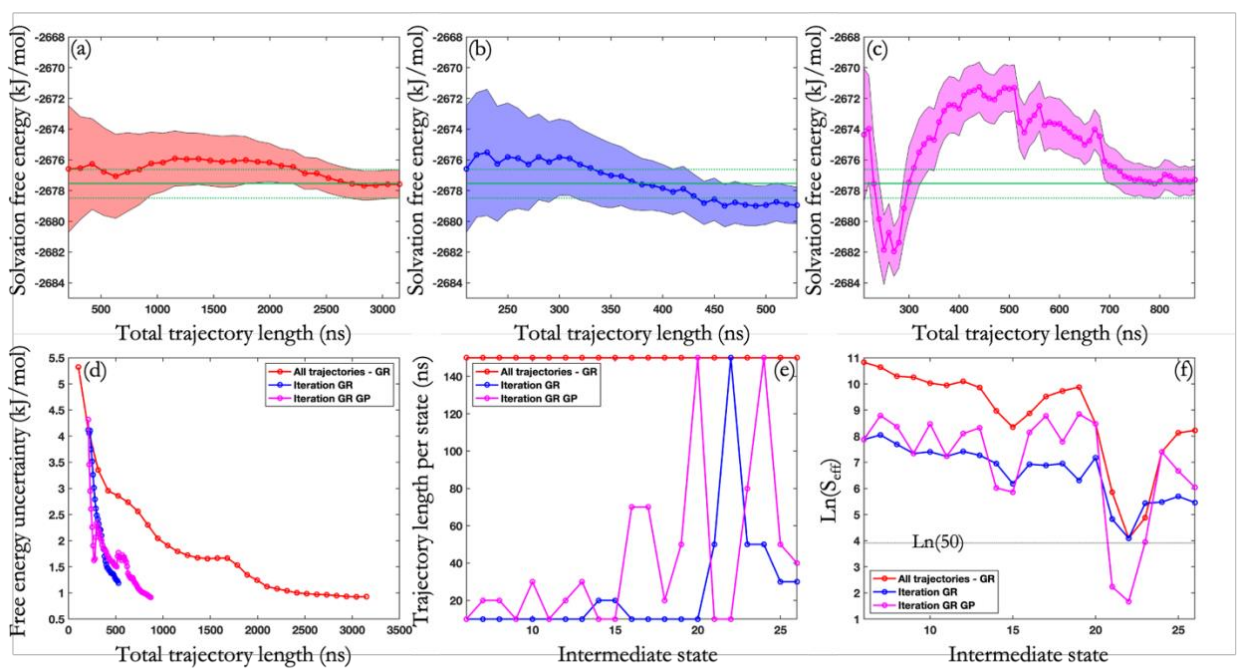

Figure S6. Evolution of Coulombic contribution to solvation free energy at 600 K. MAP was used to determine values for the hyperparameters in GP regressions.

For Algorithm II with GP regression-MAP and GR diagnostics, while energy and uncertainty appear to converge during iterations 20 to 30 (i.e., 410 ~ 510 ns total trajectory length), four close neighboring intermediate states (e.g., 21, 22, 24, and 26) have an $S_{eff}$ below the threshold of 50 (see Table S2). This observation necessitates the undertaking of additional iterations and prolonged trajectories to ensure convergence to accurate values. Beyond approximately 55 iterations (i.e., 760 ns total trajectory length), only two intermediate states (e.g., 21 and 22) retain an $S_{eff}$ value below 50, as shown in Figure S6f.

Table S2. The $S_{eff}$ of intermediate states 21 to 26 at iterative step 25 and 65.

| Iterative step | IS 21 | IS 22 | IS 23 | IS 24 | IS 25 | IS 26 |
|---|---|---|---|---|---|---|
| 25 | 9.35 | 5.32 | 51.97 | 8.13 | 785.42 | 21.84 |
| 65 | 9.35 | 5.32 | 51.97 | 1444.62 | 785.42 | 420.06 |

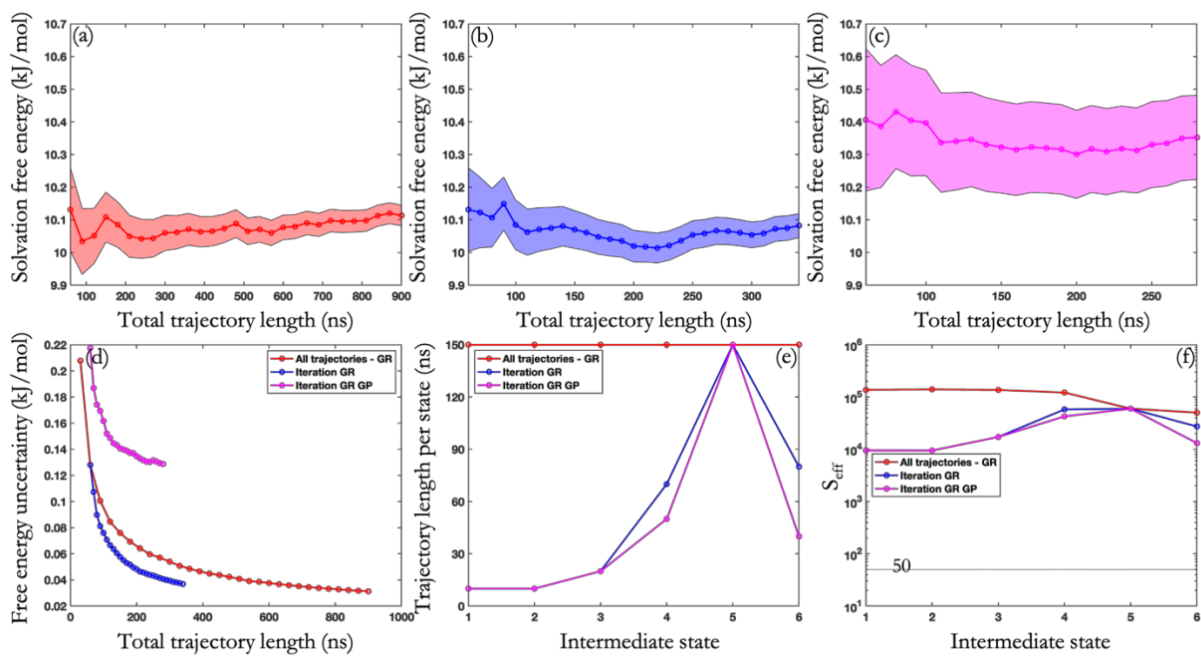

Figure S7. Evolution of van der Waals contribution to solvation free energy at 600 K using (a) TI method with GR diagnostics by uniformly expanding trajectories at each intermediate state, (b) Algorithm I (accelerated TI with GR diagnostics), and (c) Algorithm II (accelerated TI with GP regression-MAP and GR diagnostics). (d) Evolution of solvation free energy uncertainties corresponding to (a-c). (e) Trajectory length and (f) $S_{eff}$ for each intermediate state at the end of the calculation corresponding to (a-c).

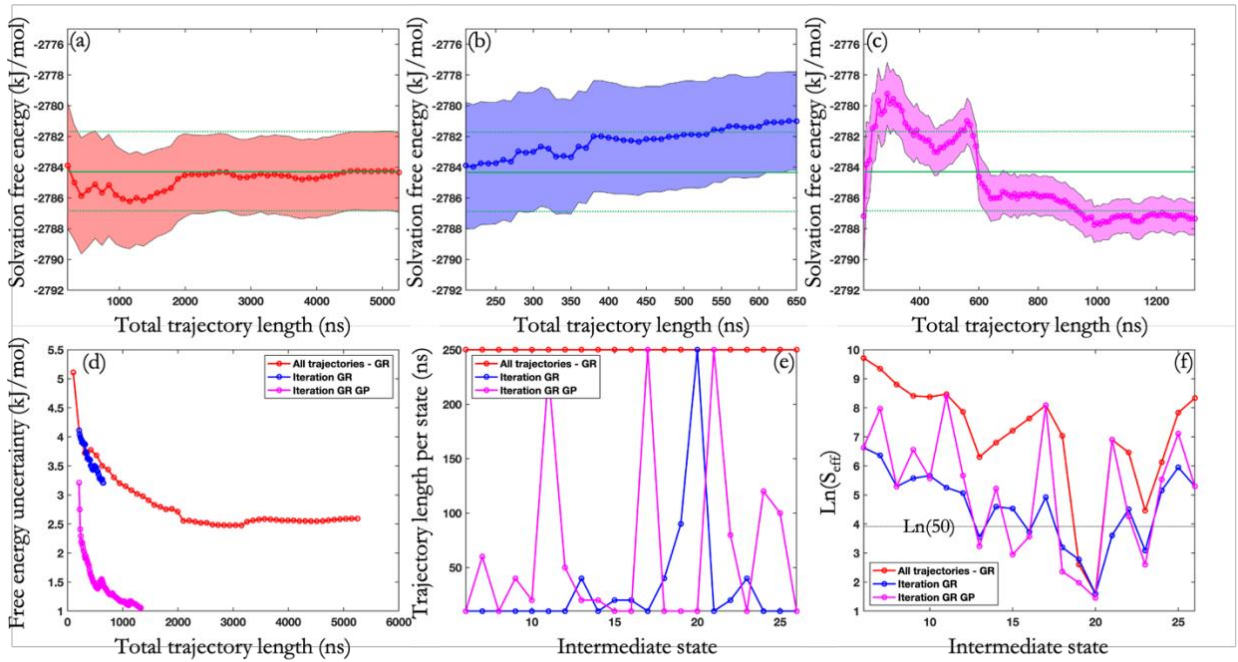

Figure S8. Evolution of Coulombic contribution to solvation free energy at 400 K. MAP was used to determine values for the hyperparameters in GP regressions.

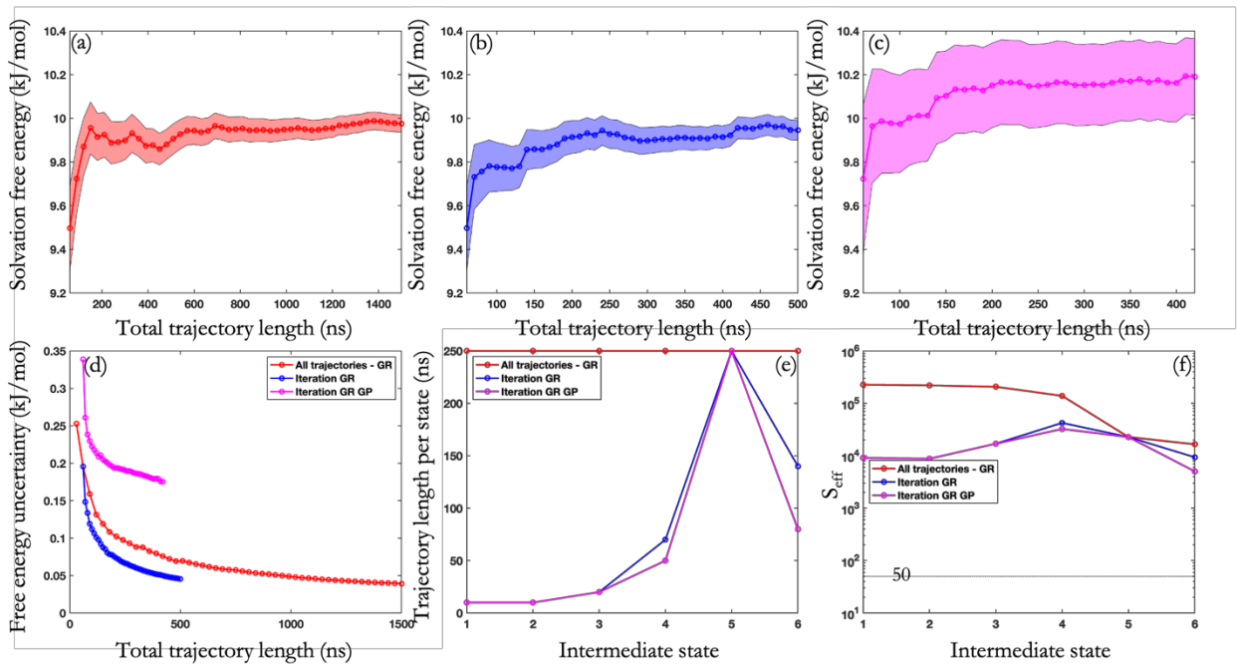

Figure S9. Evolution of van der Waals's contribution to solvation free energy at 400 K using (a) TI method with GR diagnostics by uniformly expanding trajectories at each intermediate state, (b) Algorithm I (accelerated TI with GR diagnostics), and (c) Algorithm II (accelerated TI with GP regression-MAP and GR diagnostics). (d) Evolution of solvation free energy uncertainties corresponding to (a-c). (e) Trajectory length and (f) ln ($S_{eff}$) for each intermediate state at the end of the calculation corresponding to (a-c).

Discussion on the van der Waals's contribution to solvation free energy (Figure S7 and S9): Firstly, it is noteworthy that van der Waals interactions make a substantially smaller contribution to the solvation free energy and its associated uncertainty compared to Coulombic interactions. The number of uncorrelated samples ($S_{eff}$) for all six intermediate states, at both 600 K and 400 K, significantly surpasses the empirical convergence threshold. This observation underscores that van der Waals contributions to solvation free energy can be reliably converged. Meanwhile, we found that van der Waals contributions to the solvation free energy and associated uncertainties calculated using Algorithm I are close to those obtained through the traditional TI method with GR diagnostics, employing either 150 ns or 250 concatenated trajectories for each intermediate state at 600 K or 400 K, respectively. Furthermore, we found the computed van der Waals contributions to solvation free energy and uncertainties using Algorithm II exhibit a slight discrepancy when compared to Algorithm I. This discrepancy can be attributed to the inherent limitation of numerical integration in the TI method (i.e., trapezoidal rule), stemming from the use of a limited number of data points (i.e., 6 intermediate states for van der Waals interactions). Nevertheless, it is imperative to emphasize that this energy difference remains below 0.25 kJ/mol at both temperatures, attesting to the reliability of our newly developed algorithms. In addition to their accuracy, these innovative approaches offer a substantial reduction in computational costs, with the two newly developed algorithms requiring only approximately ~30% of the computational resources compared to the TI method with GR diagnostics when employing all available trajectories for each intermediate state at 600 K or 400 K.